\documentclass[aps,prd,preprint,nofootinbib]{revtex4}
\usepackage{amsfonts}
\usepackage{mathrsfs}
\usepackage{graphicx}
\usepackage{amsmath}
\usepackage{amssymb}
\usepackage{epsfig}
\usepackage{graphicx}
\usepackage{slashed}
\usepackage{color}
\usepackage{multirow}
\usepackage{ulem}
\usepackage{adjustbox}

\usepackage{stmaryrd}
\newcommand{\bqa}{\begin{eqnarray}}
	\newcommand{\eqa}{\end{eqnarray}}
\newcommand{\beq}{\begin{equation}}
	\newcommand{\eeq}{\end{equation}}
\allowdisplaybreaks[1]

\graphicspath{{fig/}{dia/}} \DeclareGraphicsExtensions{.eps}
\begin{document}

\title{Time-reversal asymmetries  in $\Lambda_b$  semileptonic decays }

\author {Chao-Qiang Geng\footnote{cqgeng@ucas.ac.cn}, Xiang-Nan Jin\footnote{jinxiangnan21@mails.ucas.ac.cn} and  Chia-Wei Liu\footnote{chiaweiliu@ucas.ac.cn}}
\affiliation{	
	School of Fundamental Physics and Mathematical Sciences, Hangzhou Institute for Advanced Study, UCAS, Hangzhou 310024, China\\
	University of Chinese Academy of Sciences, 100190 Beijing, China
}
\date{\today}

\begin{abstract}
	We study the decays of $\Lambda_b \to \Lambda_c(\to B_n f) \ell ^- \overline{\nu}$ with $\ell = e, \mu, \tau$, where $B_n$ 
	and $f$ are the daughter baryon and  the rest of the particles in $\Lambda_c$ cascade decays, respectively.
	In particular, we examine  the full angular distributions with polarized $\Lambda_b$ and  lepton mass effects, in which 
	the time-reversal asymmetries are identified.
		We  concentrate on the decay modes of $\Lambda_b \to \Lambda_c(\to p K^- \pi^+) \ell ^- \overline{\nu}$ 
	to demonstrate their experimental feasibility. 
	We show that the  observables associated  with the time-reversal asymmetries are useful  to  search for new physics 
	as they vanish in the standard model. 
We find that they are sensitive to  the right-handed current from new physics, and possible to be observed at LHCb. 	
\end{abstract}

\maketitle

\section{Introductions}
The transitions of $b\to c \ell ^- \overline{\nu}$  with $\ell = e , \mu , \tau$ have raised great interest in both the theoretical and experimental aspects~\cite{Bmeson,BmesonEXP}. 
In particular, 
the discrepancy of   $R_{D^{(*)}} = \Gamma ( \bar {B} \to D^{(*)} \tau^- \overline{\nu }) / \Gamma ( \bar {B} \to D^{(*)} l ^- \overline{\nu }) $ with $l = e, \mu $
has shown  that the possible contributions from new physics~(NP) can be as large as ${\cal O}(10 \%)$.
Explicitly, we have that $R_{D,D^*}^{\text{exp}} = (0.340 \pm0.030 ,  0.295 \pm 0.014)$ 
from the experiments~\cite{BmesonEXP,pdg,HFLAV:2019otj}  and $R_{D,D^*}^{\text{SM}} = (0.304 \pm0.003 ,  0.259 \pm 0.006)$ 
from  the lattice QCD calculation~\cite{latticeBmeson}, implying  that NP can play a significant role. 
For a review, one is referred to Ref.~\cite{ReVonLUV}. 

On the other hand, LHCb has recently announced the baryonic version of the ratio 
to be $R_{\Lambda_c} = \Gamma ( \Lambda_b \to \Lambda_c \tau ^ - \overline{\nu }) /  \Gamma ( \Lambda_b \to \Lambda_c l ^ - \overline{\nu })  = 0.242 \pm0.026\pm0.040\pm0.059$~\cite{LbMeasurements tau }, where 
the first and second uncertainties are statistical and systematic, respectively, and 
the third one comes from the normalization channel of $\Lambda_b \to \Lambda_c \pi^+ 2\pi^-$. 
In contrast to $R_{D^{(*)}}$,   $R_{\Lambda_c}$ is found to be larger in theory, given as $R_{\Lambda_c}= 0.324\pm 0.004 $ based on lattice QCD~\cite{lattice}.
Such opposite behavior indicates that there would be some  theoretical errors, which have not been  properly considered.
Thus, as a complementarity, it is useful to examine the angular distributions~\cite{Unpolarized,Korner,Tau Decay}. 
In most of the works in the literature, $\Lambda_b$ is assumed to be unpolarized. However,
it is important to analyze the polarized cases, since the  polarization fraction $P_b$ is recently found to be around $3\%$ in proton-proton collisions at center-of-mass energies of  13 TeV~\cite{polarizedEXP}. 
We emphasize that with $P_b\neq 0$,  the time-reversal~(TR) asymmetries  can be observed  without the cascade decays of  $\Lambda_c$ as we will show in this work. Moreover,  the value of $3\%$ is twice larger than ${\cal B}(\Lambda_c \to \Lambda \pi^+ , pK_S^0) $, and hence  it is useful to study the cases with $P_b\neq 0 $ for probing the TR asymmetries.

The angular distribution of $\Lambda_b \to \Lambda_c ( \to pK_S^0 )\mu ^- \overline{\nu}$ with polarized $\Lambda_b$  was  first given in Ref.~\cite{polarized}.
In this work, we provide the full angular distributions of $\Lambda_b \to \Lambda_c(\to B_n f) \ell^- \overline{\nu}$, where $B_n$ is the daughter baryon and $f$ stands for the rest of the daughter particles. 
In contrast to those in the literature,  we extend the study to  the three-body $\Lambda_c$ decays  to include $\Lambda_c \to p K^- \pi ^+$ and
$\Lambda_c \to \Lambda l^+ \nu$.  In particular, $\Lambda_c \to p K^- \pi^+$  has a great advantage for the experimental detection, since all the particles  in  the final states are charged.

In the standard model~(SM), the TR asymmetries in  $\Lambda_b \to \Lambda_c(\to B_n f) \ell^- \overline{\nu}$ are zero due to the absence of the weak phase
in the $\Lambda_b \to \Lambda_c$ transition. Clearly, a nonvanishing TR asymmetry indicates the existence of NP with a new {\it CP} violating phase beyond the SM.

The layout of this work is given as follows. In Sec.~\MakeUppercase{\romannumeral 2},  we present the angular distributions of the SM parametrized by the helicity amplitudes. 
In Sec.~\MakeUppercase{\romannumeral 3},
we discuss the effects from NP, and show that they can be absorbed by redefining the helicity amplitudes.  In Sec.~\MakeUppercase{\romannumeral 4},  we estimate the TR asymmetries and their feasibility to be measured at LHCb. At last, we conclude the study in  Sec.~\MakeUppercase{\romannumeral 5}.

\section{Decay observables}
In the SM, 
the amplitudes of  $ \Lambda_b \to  \Lambda_c \ell^- \overline{\nu}$ are dominated by the weak interaction at  tree level, given as 
\begin{eqnarray}\label{eq1}
&&\frac{G_F}{\sqrt 2} V_{cb}  g ^{\mu \nu}    \overline{u}_{\ell} {\gamma_{\mu}} (1-{\gamma_5}) v   \langle \Lambda _c | \overline{c}{\gamma_{\nu}}(1-{\gamma_5})b | \Lambda _b \rangle,
\end{eqnarray}
where $G_F$  is the Fermi constant,  $V_{cb}$ corresponds to the Cabibbo-Kobayashi-Maskawa~(CKM) matrix element, and $u_\ell $ and $v$ are the Dirac spinors of  charged leptons and  antineutrinos, respectively. In this work, we do not specify the  flavors of  (anti)neutrinos as they cannot be distinguished in the experiments.

We  further decompose  the amplitudes by expanding the Minkowski metric,
\begin{equation}\label{eq2}
g^{\mu\nu} = \varepsilon_t ^{\mu } (q)\varepsilon_t^{\ast \nu} (q)- \sum_{\lambda = 0,\pm } \varepsilon^\mu _\lambda(q) \varepsilon^{\ast \nu}  _ \lambda (q) \,,  
\end{equation}
where $ q=(q^0,\vec{q}\,)$ and   $\varepsilon$ are the four-momentum and   polarization vector of the off-shell $W$ boson, respectively.  The  subscript in $\varepsilon$ denotes  the helicity, where $t$ indicates  timelike while the others  spacelike. 
In particular, we  have  that
\begin{equation}\label{restPolar}
	{\varepsilon} ^\mu _\pm  = \frac{1}{\sqrt{2}}(0,  \pm 1, i , 0 )^T\,, \quad {\varepsilon}_0^\mu  = (0 ,0,0,-1 )^T \,, \quad  {\varepsilon}^\mu  _t = ( - 1 ,0,0,0 )^T\,,
\end{equation}
in the center of mass frame of $\ell^-\overline{\nu}$, which would be referred  to  as the $\vec{q}$ frame in the following.  
Notice that the relative phases between $\varepsilon$ are crucial as they interfere in the decay distributions. In this work, they are fixed by the lowering operators, given by 
\begin{equation}
\left( J_x -  i J_y\right)  \varepsilon_{1,0}  = \sqrt{2 }\varepsilon_{0,-1}\,, 
\end{equation}
where $J_{x,y}$ are the $SO(3)$ rotational generators. 
On the other hand, in the center of the mass frame of  $\Lambda_b$ with $\vec{q} = -|\vec{q}\,|\hat{z}$, which would be  referred to as  the $\Lambda_b$ frame, we have 
\begin{equation}\label{qframe} 
	{\varepsilon} ^\mu _\pm  = \frac{1}{\sqrt{2}}(0,  \mp 1, i , 0 )^T\,, \quad {\varepsilon}_0^\mu  = \frac{1}{\sqrt{q^2}}(-|\vec{q}\,| ,0,0,q^0 )^T \,, \quad  {\varepsilon}^\mu  _t =\frac{-1}{\sqrt{q^2}} q ^\mu ,
\end{equation}
which are useful for the latter purpose. 

Plugging Eq.~\eqref{eq2} in Eq.~\eqref{eq1}, we have 
\begin{equation}\label{good}
\frac{	G_F }{\sqrt{2}} V_{cb} \left(
L_t B_t   - \sum_{\lambda=0,\pm}  L_\lambda  B_\lambda 
\right)\,,
\end{equation}
and 
\begin{equation}\label{good2}
B_{\lambda_W}   =  \varepsilon_{\lambda_W}  ^{\ast\mu} \langle \Lambda _c | \overline{c}{\gamma_ {\mu}}(1-{\gamma_5})b | \Lambda_b \rangle\,,\quad  L_{\lambda_W}  =  \varepsilon_{\lambda_W} ^{\mu} \overline{u}_{\ell} {\gamma_ {\mu}} (1-{\gamma_5}) v \,,
\end{equation}
with  ${\lambda_W}  = t, 0$ and  $\pm$. 
Note that $B_{\lambda_W}$ and $L_{\lambda_W}$  depend on the polarizations of the baryons and leptons, respectively. 
It is clear that in  Eqs.~\eqref{good} and \eqref{good2},
 the amplitudes are decomposed as the products of Lorentz scalars, describing $\Lambda _b \to \Lambda _c W^{-*}$~($B_{\lambda_W}$) and $W^{-*} \to \ell^- \overline{\nu}$~($L_{\lambda_W}$).   
A great advantage is that $B_{\lambda_W}$ and $L_{\lambda_W}$  can be computed 
 independently in the  $\Lambda_b$  and $\vec{q}$ frames, respectively,  reducing the three-body problems to the  products  of   two-body ones.

To proceed further, we have to consider the polarizations of the baryons and leptons.
To this end, it is convenient to parametrize $B_{\lambda_W}$ as 
\begin{eqnarray}\label{number45}
B_{\lambda_W} =\varepsilon^{\ast\mu} _{\lambda_W} \overline{u}_{c}&&\left[ \left(f _1(q^2) {\gamma_{\mu}} - i f _2(q^2) \frac{\sigma_{\mu \nu}}{M_b}q^{\nu} + f _3(q^2)\frac{q_{\mu}}{M_b} \right)\right.  \nonumber\\
&&~~~- \left. \left(g _1(q^2) {\gamma_{\mu}} -ig  _2(q^2) \frac{\sigma_{\mu \nu}}{M_b}q^{\nu} + g _3(q^2)\frac{q_{\mu}}{M_b} \right) {\gamma_5} \right]  u_{b},
\end{eqnarray}
where $f_{1,2,3}$ and $g_{1,2,3}$  represent the form factors,  $M_{b}$ is the mass of $\Lambda _{b}$, and $\sigma_{\mu \nu} = i(\gamma_\mu \gamma_{\nu} - \gamma_\nu \gamma_\mu)/2$. 
The helicity amplitudes are calculated by 
\begin{eqnarray}
H_{\lambda_c, \lambda_W} 
= B_{\lambda_W}\left(
\lambda_b  = \lambda_c - \lambda_W, \lambda_c , \vec{p}_c = -\vec{q} = |\vec{p}_c | \hat{z}
\right)
\,,
\end{eqnarray}
where $\lambda_{b(c)} $ corresponds to  the angular momentum~(helicity) of $\Lambda_{b(c) }$,
$\vec{p}_c$ is the three-momentum of $\Lambda_c$ in  the  $\Lambda_ b$ frame, and the conventions of the Dirac spinors are given in Appendix A. 
Plugging Eq.~\eqref{qframe} in Eq.~\eqref{number45}, we obtain explicitly  that
\begin{eqnarray}\label{helicity}
&&	H_{\pm \frac{1}{2} \pm  1}=\sqrt{2 Q_{-}}\left(  f_1  +  \frac{M_{+}}{M_{b}} f_{2}\right)
\pm  \sqrt{2 Q_{+}}\left(-g_1+ \frac{M_{-}}{M_{b}}g_2 \right)	\,,\nonumber\\
&&	H_{\pm \frac{1}{2} 0} = - \sqrt{ \frac{Q_{-}}{q^{2}}} \left(M_{+} f_1 + \frac{q^{2}}{M_{b}} f_2 \right)\pm  
	\sqrt{ \frac{Q_{+}}{q^{2}}}\left(M_{-} g_1 - \frac{q^{2}}{M_{b}} g_2 \right)\,,
	\nonumber\\
&&	H_{\pm \frac{1}{2} t}=- \sqrt{ \frac{Q_{+}}{q^{2}}}\left(M_{-} f_1 + \frac{q^{2}}{M_{b}} f_3  \right)\pm
\sqrt{ \frac{Q_{-}}{q^{2}}}\left(M_{+} g_1 - \frac{q^{2}}{M_{b}} g_3 \right)\,, 
\end{eqnarray}
where  
 $M_\pm = M_{b } \pm 
M_{c}$, $M_c$ is the mass of $\Lambda_c$,  and $Q_\pm = (M_\pm )^2 - q^2  $. 
Note that both
the form factors and  amplitudes depend on  $q^2$. 

On the other hand, the antineutrinos have positive helicities, and  $L_{\lambda_W} $ depends only on $\lambda_\ell$ the helicity of $\ell^-$.  
From the definitions of $h_{\pm}$, given by
\begin{eqnarray}
&&	h_+ = L_0  \left (\lambda_e = \frac{1}{2}, \vec{p}_\ell = -\vec{p}_\nu = |\vec{p}_e| \hat{z}\right )\,, \nonumber\\
&&	h_- = L_{-1}  \left (\lambda_e = -\frac{1}{2}, \vec{p}_\ell = -\vec{p}_\nu = |\vec{p}_e| \hat{z}\right )\,,
\end{eqnarray}
we explicitly have 
\begin{equation}\label{ratiosof Lepton}
h_- = -  2 \sqrt{2 (q^2 - m_{\ell} ^2) } \,,~~~
h _+  
= \sqrt{  \delta_\ell} h_ - \,,~~~
 \delta_\ell = \frac{m_\ell^2 }{2 q^2 }\,,
\end{equation}
with   Eqs.~\eqref{restPolar} and \eqref{good2}
with
$\vec{p}_{\ell(\nu) }$  the three-momentum of  $\ell^-(\overline{\nu})$ in the  $\vec{q}$ frame\,. 

\begin{figure}[b]
	\includegraphics[width=0.96 \linewidth]{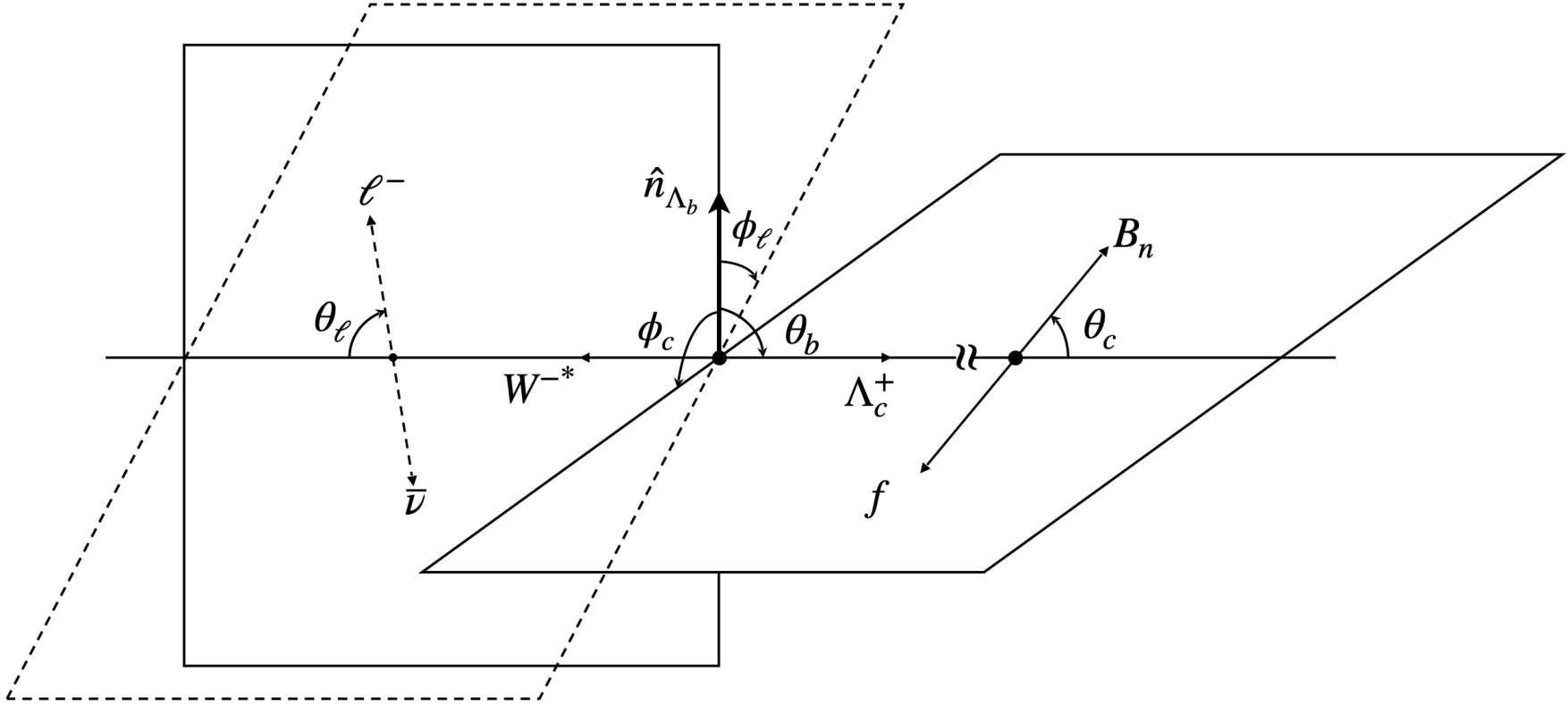}  	
	\caption{ Definitions of the angles, where $B_n$ represents the daughter baryon and $f$ the rest of the  decay particles. }
	\label{PA3} 
\end{figure}

The angular  distributions  of $\Lambda_b \to \Lambda_c ( \to B_n f) \ell ^- \overline{\nu}$
can be obtained by piling up the Wigner-d matrices of $d^J$,  read as 
\begin{eqnarray}\label{distru}
&&\frac{\partial^6\Gamma (\Lambda_b \to \Lambda_c ( \to B_n f) \ell ^- \overline{\nu}) )}{\partial q ^2 \partial \cos \theta_b \partial \cos \theta_c  \partial \cos \theta_\ell  \partial \phi_c \partial\phi_\ell  } ={\cal B}(\Lambda_c \to B_nf )  \frac{\zeta (q^2)}{32\pi ^2 } \sum_{\lambda_\ell \,,\lambda \,,\lambda_b  }\rho_{\lambda_b,\lambda_b }\left| A^c_{\lambda } h_{ \lambda_\ell }\right|^2  \nonumber\\
&&   \left|
\sum_{\lambda_c\,, \lambda_W } (-1) ^{J_W  }
H_{\lambda_c\,,\lambda_W }
d^{\frac{1}{2}}(\theta_b)^{\lambda_b }\, _{\lambda_c - \lambda_W }
d^{\frac{1}{2}}(\theta_c)^{\lambda_c}\,_{\lambda }
d^{J_W }(\theta_\ell )^{\lambda_W }\,_{\lambda_\ell  - \frac{1
	}{2}}
e^{i(\lambda_c\phi_c +\lambda_\ell  \phi_\ell )} 
\right|^2,\nonumber\\
&&\zeta (q^2)=  \frac{G_F^2 }{24 \pi^3}|V_{cb}| ^2 \frac{(q^2-M_{\ell}^2)^2|\vec{p}_c|  }{8M_{ b }^2q^2} \,, 
\end{eqnarray}
where
${\cal B}(\Lambda_c \to B_n f)$ are the branching fractions of $\Lambda_c \to B_n f$,
  $\rho_{\pm,\pm } = (1\pm P_b)/2$, 
$\lambda_{(b,c,\ell )} = \pm 1/2$ , 
$|\vec{p}_c|  = \sqrt{Q_+Q_-}/2M_{{\bf B}_b}$,  the factor of  $(-1)^{J_W}$ comes from Eq.~\eqref{good}  along with $J_W = 0~(1)$ for $\lambda_W = t~(\pm ,0)$, 
and   $A_{\lambda }^c  $ are associated with the   up-down asymmetries of $\Lambda_c\to B_n f$.  Here, the definitions of the angles can be found in FIG.~\ref{PA3}, 
where $\theta_{b,c}$ and $\theta_\ell$ are defined in  the center of mass frames of $\Lambda _{b,c}$ and $\ell^- \overline{\nu}$, respectively, while $\phi_{c,\ell}$ are the azimuthal angles between the decay planes.

The derivation of Eq.~\eqref{distru} is sketched in Appendix~\ref{APPB}. 
The index  $\lambda$ corresponds to $\lambda_{B_n}  -\lambda_f$ with $\lambda_{B_n }$ and $\lambda_f$ the helicities of $B_n$ and  $f$ in $\Lambda_c \to B_n f$, respectively. If $f$ contains more than two particles, we simply group them together,  forming an  angular momentum eigenstate in the center-of-mass frame of $f$, acquiring an effective helicity. 

 In the case of  $\Lambda_c\to p K^- \pi^+~(\Lambda l^+ \nu) $, $A_\lambda^c$   depends on the  three-momentum of  $p(\Lambda)$ and angles in $K^-\pi^+$ $(l^+\nu)$  as well. However, we integrate out the dependence for simplicity in this work. 
In addition,  
 the  cascade decays of $\tau^-$  can be included  by continually piling up the  Wiger-d matrices inside Eq.~\eqref{distru}. The interested readers are referred to Ref.~\cite{Tau Decay}.
Note that
the overall $q^2$ dependence in Eq.~\eqref{distru} can be cast in a more symmetric form by recognizing  $| \vec{p}_\ell|= (q^2- m_\ell^2) / \sqrt{4 q^2}$ in the $\vec{q}$ frame.

We  expand the angular distributions as  
\begin{align}\label{decomposed}
\begin{aligned}
&\frac{\partial^6 \Gamma (\Lambda_b \to \Lambda_c ( \to B_n f) \ell ^- \overline{\nu}) }{\partial q ^2 \partial \vec{\Omega}  } =\\
&\quad\quad{\cal B}(\Lambda_c \to B_nf )  \frac{\zeta (q^2)}{32\pi ^2 }    \sum_{i=1}^{26} \text{Re}\left( {\cal X}_i (a_\pm, b_\pm,t_\pm) {\cal D}_i (\vec{\Omega }) \right)  {\cal P}_i(\alpha,P_b) ,
\end{aligned}
\end{align}
where  $\alpha$ are the up-down asymmetries of $\Lambda_c \to B_n f$,  $\vec{\Omega} = (\theta_{b,c,\ell}\,,\phi_{c,\ell})$, and   
 explicit forms of ${\cal X}_i$, ${\cal P}_i$  and ${\cal D}_i$ can be found in Table~\ref{Table}, where we have taken the abbreviations:
\begin{equation}\label{abbriev for tab}
a_\pm = H_{\pm\frac{1
	}{2} ,0 }\,,~~~b_\pm = H_{\mp\frac{1
	}{2}, \mp 1} \,,~~~t_\pm = H_{ \pm \frac{1
}{2}, t} \,,~~~|\xi ^2| = |\xi_+^2| +|\xi_-^2|  \,,~~~|\xi _\Delta ^2| = |\xi_+^2| - |\xi_-^2|\,,
\end{equation}
with $\xi = a,b,t$
and  $P_2 = (
3 \cos^2\theta_\ell -1)/2$. 
The real-valued function in Eq.~\eqref{decomposed} guarantees that the partial decay widths are real. 
For an illustration, we have 
\begin{eqnarray}\label{equationTR}
&&\text{Re}\left({\cal X}_{7} {\cal D}_{7} \right)= 
\frac{3}{\sqrt{2}} \left[ \text{Re}\left(  2 \delta_\ell (t_-b_+^* - b_-t_+^*) + (a_-b_+^* + b_-a_+^*)\right) \cos\phi_\ell \sin \theta_b \sin \theta_\ell   -  \right.\nonumber\\
&&\quad\quad\quad\left. {\bf \text{Im}}\left(  2 \delta_\ell (t_-b_+^* - b_-t_+^*) + (a_-b_+^* + b_-a_+^*)\right) \sin \phi_\ell \sin \theta_b \sin \theta_\ell  \right] \,,
\\
&&\text{Re}\left(
{\cal X}_{12} {\cal D}_{12}\right) = 
\frac{3}{\sqrt{2}}\left[  \text{Re}\left(  2 \delta_\ell (b_-t_-^* - t_+b_+^*) -  (a_+b_+^* + b_-a_-^*)\right) \cos(\phi_\ell +\phi_c) \sin \theta_c \cos \theta_c    \right. \nonumber\\
&&\quad\quad\quad-\left. {\bf \text{Im}}\left(  2 \delta_\ell (b_-t_-^* - t_+b_+^*) -  (a_+b_+^* + b_-a_-^*)\right) \sin (\phi_\ell +\phi_c)  \sin \theta_c \cos \theta_c  \right] \,,
\end{eqnarray}
by the  identity of
	$\text{Re}( {\cal X}_i{\cal D}_i  )=  \text{Re}({\cal X}_i)\text{Re}({\cal D}_i) -  {\bf \text{Im}}({\cal X}_i){\bf \text{Im}}({\cal D}_i)$. 
For those ${\cal D}_i $, which are independent of $\phi_{c,\ell}$, we simply have Re$({\cal X}_i {\cal D}_i)= {\cal X}_i{\cal D}_i $. 
Notice that $\xi_\pm $ are real  in the SM, and any observations of nonzero ${\bf \text{Im}}( {\cal X}_i)$   would be  a smoking gun of NP. 
The angular distributions of $\overline{\Lambda_b} \to \overline{\Lambda_c}(\to \overline{ B_n }  \bar {f } ) \ell^+ \nu$ can  be obtained directly by taking $\theta_\ell \to \pi - \theta_\ell$ and $\alpha\to -\alpha$.  
In practice, $\delta_{l}$ can be taken as zero as an excellent approximation in the SM, with which $t_\pm$ can be neglected as well since they are always followed by  $\delta_\ell$.

It is interesting to point out that, under the parity transformation, the helicity amplitudes behave differently as 
\begin{equation}
a_\pm \to a_\mp\,,~~~b_\pm \to b_\mp\,,~~~t_\pm \to - t_\mp\,,
\end{equation}
so that Re$({\cal X}_{7,12})$ and  Im$({\cal X}_{7,12})$ are parity even and odd, respectively. 
If $\Lambda_b$ is unpolarized~($P_b=0$), it is clear that  $\phi_c$ and $\phi_\ell$ can not be measured separately. In this case,  it is convenient to introduce a new set of azimuthal coordinates as 
\begin{equation}
	\begin{aligned}
	&\Phi=\phi_\ell  + \phi_c \,,\,\,\,\,\,\,\,\,\,&0<\Phi  <2\pi\,, \\
&\Phi_\Delta = \frac{1}{2}(\phi_\ell  -\phi_c)\,,\,\,\,\,\,\,&-\pi <\Phi_\Delta <\pi\,.
	\end{aligned}
\end{equation}
To obtain the  unpolarized angular distributions from the polarized ones, one can integrate over $\Phi_\Delta$ and $\cos \theta_b$, in which  ${\cal D}_{4\text{--}8,14\text{--}26 }$ are  zero. As  a cross-check, we find that the results are identical to those given in Ref.~\cite{Korner}.

\begin{table}[hbt!]
	\caption{The  angular distributions of $\Lambda_b \to \Lambda_c(\to B_n f ) \ell ^- \overline{\nu}$ with   	$P_2 = (3\cos^2 \theta_\ell -1)/2$ and the parameters of 
 $a$, $b$  and  $t$ defined in Eq.~\eqref{abbriev for tab}. } \label{Table}
\resizebox{0.78\textwidth}{!}{
	\begin{tabular}{l|ccc}
		\hline
		$i $&${\cal X}_i$ &
		$P_i$ & ${\cal D}_i$ 	\\
		\hline
		1&$ (\delta_\ell + 1)(|a|^2 + |b|^2) + 3\delta_\ell |t|^2 $&$1$ &$ 1 $ \\
		2&$ (2\delta_\ell-1)(|a|^2 - \frac{1}{2} |b|^2) $&$1$ &$P_2$\\
		3&$ - 6\delta_\ell\left(\text{Re}\left( a_+t_+^*\right) +\text{Re} \left( a_-t_-^*\right) \right )+ \frac{3}{2}|b_\Delta|^2$&$1$ &$ \cos \theta_\ell$ \\
		4&$ (\delta_\ell + 1)(|a_\Delta|^2 + |b_\Delta|^2) + 3\delta_\ell |t_\Delta|^2$&$P_b$&$ \cos \theta_b $\\ 
5 &$(2\delta_\ell-1)(|a_\Delta|^2 - \frac{1}{2}|b_\Delta|^2)$&$P_b$&$ \cos \theta_b P_2$\\ 
6&$- 6\delta_\ell \left[\text{Re}(a_+t_+^*) - \text{Re}(a_-t_-^*)\right] - \frac{3}{2}|b_\Delta|^2$&$P_b$&$ \cos \theta_b \cos \theta_\ell  $\\
7&$\frac{3}{\sqrt{2}}\left[ 2 \delta_\ell (t_-b_+^* - b_-t_+^*) + (a_-b_+^* + b_-a_+^*)\right]$&$P_b$&$ e^{i \phi_{\ell}}\sin \theta_b \sin \theta_\ell $\\
8&$ \frac{3}{\sqrt{2}} (2\delta_\ell- 1) (b_-a_+^* - a_-b_+^*) $&$P_b$&$ e^{i \phi_{\ell}} \sin \theta_b \sin \theta_\ell \cos \theta_\ell$\\
		9&$ (\delta_\ell + 1)(|a_\Delta|^2 - |b_\Delta|^2) + 3\delta_\ell |t_\Delta|^2$&$\alpha$&$ \cos \theta_c $\\
		10&$ (2\delta_\ell-1)(|a_\Delta|^2 + \frac{1}{2}|b_\Delta|^2)$&$\alpha$ &$ \cos \theta_c P_2$ \\
		11&$6\delta_\ell\left( \text{Re} \left( a_-t_-^*\right) - \text{Re}\left( a_+t_+^*\right)  \right ) - \frac{3}{2}|b|^2 $&$\alpha$&$ \cos \theta_c \cos \theta_\ell $\\
		12&$ \frac{3 }{\sqrt{2}}\left[ 2 \delta_\ell (b_-t_-^* - t_+b_+^*) -  (a_+b_+^* + b_-a_-^*)\right]$&$\alpha$&$ e^{i\Phi} \sin \theta_c \cos \theta_c $\\
		13&$ \frac{3}{\sqrt{2}}\left[ 2 \delta_\ell (a_+b_+^*- b_-a_-^* ) + (b_-a_-^*- a_+b_+^* )\right]$ &$\alpha$ &$ e^{i\Phi} \sin \theta_c \cos \theta_c \cos \theta_\ell$\\
		14&$ (2\delta_\ell+1)a_+ a_-^* $&$P_b \alpha$&$  \sin \theta_b \sin \theta_c P_2$\\
		15&$ 2\delta_\ell (|a|^2 + |b|^2)-(|a|^2 + 1/2|b|^2) $&$P_b \alpha$&$ \cos \theta _b \cos \theta_c P_2$\\		
		16&$ \delta_\ell(|a|^2 - |b|^2 + 3|t|^2 )+(|a|^2 - |b|^2)$&$P_b \alpha$&$ \cos \theta_b \cos \theta_c$\\
		17&$  -6\delta_\ell \left( \text{Re}(a_+t_+^*) +\text{Re}(a_-t_-^*)\right)  - \frac{3}{2}|b_\Delta|^2$&$P_b \alpha$&$  \cos \theta_b \cos \theta_c \cos \theta_\ell$\\
			18&$ (2-4\delta_\ell) a_- a_+^*$&$P_b \alpha$&$ e^{-i \phi_c} \sin \theta _b \sin \theta_c P_2$\\
	19 &$ (1-2\delta_\ell) b_- b_+^*$&$P_b \alpha$&$ e^{i(\phi_c + 2\phi_{\ell})} \sin \theta_b \sin \theta_c P_2$\\
		20&$ -2\delta_\ell( a_+ a_-^* + 3t_+ t_-^*)-2a_+ a_-^*$&$P_b \alpha$&$  e^{i \phi_c} \sin \theta_b \sin \theta_c$\\
		21&$ (2\delta_\ell-1)b_-b_+^* $&$P_b \alpha $&$ e^{i(\phi_c + 2\phi_{\ell})} \sin \theta_b \sin \theta_c $\\
		22&$ 6 \delta_\ell(a_+t_-^* + t_+a_-^*)$&$P_b \alpha $&$ e^{i \phi_c} \sin \theta _b\sin \theta_c \cos \theta_\ell$\\
		23&$ \frac{3}{\sqrt{2}} \left[ (b_-a_+^* - a_-b_+^*)-2\delta_\ell (b_-t_+^* + t_-b_+^*)\right]$&$P_b \alpha $&$ e^{i \phi_{\ell}} \sin \theta_b \sin \theta_\ell \cos \theta_c$\\
		24&$ \frac{3}{\sqrt{2}} \left[ (b_-a_-^* - a_+b_+^*)-2 \delta_\ell (b_-t_-^* + t_+b_+^*)\right]$&$P_b \alpha $&$ e^{i\Phi} \sin \theta_c \sin \theta_\ell \cos \theta_b $\\
		25&$ \frac{3}{\sqrt{2}}\left[2 \delta_\ell (a_-b_+^* + b_-a_+^*) - (a_-b_+^* + b_-a_+^*)\right]$&$P_b \alpha $&$ e^{i \phi_{\ell}} \sin \theta_b \sin \theta_\ell \cos \theta_c \cos \theta_\ell$\\		
		26&$ \frac{3}{\sqrt{2}} \left[2 \delta_\ell (b_-a_-^*- a_+b_+^* ) - ( a_+b_+^* + b_-a_-^*)\right]$&$P_b \alpha $&$ e^{i\Phi} \sin \theta_c \sin \theta_\ell \cos \theta_b \cos \theta_\ell$\\
		\hline
	\end{tabular}} 
\end{table}

With Table~\ref{Table},
one can  construct several observables in a model independent way.  The simplest ones would be the partial and total decay widths, read as
\begin{eqnarray}\label{number3}
&& \frac{d \Gamma (\Lambda_b \to \Lambda_c  \ell ^- \overline{\nu}) }{d q^2}= \zeta(q^2){\cal X}_1\,, 
\end{eqnarray}
and 
\begin{eqnarray}\label{number4}
	&& 
\Gamma = 	\Gamma(\Lambda_b \to \Lambda_c \ell ^- \overline{\nu})
	= \int^{(M_b -M_c)^2}_{m_\ell^2} \zeta (q^2) {\cal X}_1 dq^2  \,,
\end{eqnarray}
respectively. 
It shall be clear that  $\Gamma$ is independent of $\Lambda_c\to B_n f$.
Likewise, there are several  observables that can be defined independent of $\Lambda_c\to B_nf$, and it is reasonable to measure them separately as they do not suffer from the smallness of ${\cal B}(\Lambda_c \to B_n f )$. 
In fact, the angular distributions without cascade decays can be obtained straightforwardly by  integrating over $(\cos \theta_c,\phi_c) $, resulting in 
\begin{align}\label{alpha0}
	\begin{aligned}
		\frac{\partial^4 \Gamma(\Lambda_b \to \Lambda_c \ell ^- \overline{\nu}) }{\partial q ^2 \partial \cos\theta_b \cos \theta _\ell \partial \phi_\ell   } =\frac{\zeta(q^2)}{8\pi }  \sum_{i=1}^{8} \text{Re}\left( {\cal X}_i (a_\pm, b_\pm,t_\pm) {\cal D}_i (\vec{\Omega }) \right)  {\cal P}_i(0,P_b)\,,
	\end{aligned}
\end{align}
which is clearly independent of $\alpha$. As a cross-check, we find that Eq.~\eqref{alpha0}  reduces  to the ones  given by Ref.~\cite{KornerNeutron} with  an appropriate substitution.

There are some quantities that deserve  a closer look.
The   forward-backward asymmetries  for $W^{-*}\to l^-\overline{\nu}$ and $\Lambda_c\to B_n f$ are   defined as  
\begin{eqnarray}\label{FB}
A_{FB} =2 \left(  \int ^1_0 - \int^0 _{-1} \right) \Gamma_{\cos \theta_\ell }  d\cos \theta_\ell  = \frac{1}{\Gamma} \int^{(M_b -M_c)^2}_{m_\ell^2}  \zeta (q^2) {\cal X}_3 dq^2  \,,\nonumber\\
	A_{PL} =  \frac{2}{\alpha } \left(  \int ^1_0 - \int^0 _{-1} \right) \Gamma_{\cos \theta_c }  d\cos \theta_c  = \frac{1}{\Gamma}  \int^{(M_b -M_c)^2}_{m_\ell^2}  \zeta (q^2)   {\cal X}_9 dq^2 \,,
\end{eqnarray}
where we have adopted the shorthand  notation,
\begin{equation}
\Gamma_{\vec{\Omega}}  = \frac{1}{ {\cal B}(\Lambda_c \to B_n f ) }
\frac{1}{\Gamma}
\frac{\partial \Gamma (\Lambda_b \to \Lambda_c ( \to B_n f) \ell ^- \overline{\nu})  }{\partial {\vec{\Omega}} }\,. 
\end{equation}
The up-down asymmetries $A_{UD}$, on the other hand, are given by 
\begin{equation}\label{UD}
A_{UD} =  \frac{2}{P_b} \left(  \int ^1_0 - \int^0 _{-1} \right) \Gamma_{\cos \theta_b }  d\cos \theta_b  = \frac{1}{\Gamma}  \int^{(M_b -M_c)^2}_{m_\ell^2}  \zeta (q^2)   {\cal X}_4 dq^2 \,,
\end{equation}
which 
 require $P_b\neq 0 $  for an experimental measurement.

Here, it is an appropriate place to revisit $A_\lambda^c$ in Eq.~\eqref{distru} explicitly. We have 
\begin{equation}\label{cUP}
\left|A_{\pm\frac{1}{2}} ^c \right| ^2  = \frac{1}{2}\left(
1 \pm \alpha 
\right)\,,
\end{equation}
where $\alpha$ are the up-down asymmetries  of  $\Lambda_c\to B_n f$, with the experimental values given by~\cite{pdg}  
\begin{align}\label{alphap}
	\begin{aligned}
\alpha (\Lambda_c\to \Lambda \pi^+, &\Sigma^0 \pi^+, \Sigma^+ \pi^0, p K^0_s) =\\
& ( -0.84 \pm 0.09, -0.55\pm 0.11, -0.73 \pm 0.18 , 0.2\pm 0.5  ),
	\end{aligned}
\end{align} 
Similarly, for the three-body $\Lambda_c$ decays, we have $\alpha = A_{UD}^c$, where
  $A_{UD}^c$   are   defined by substituting $\Lambda_c\to B_n f $ for $\Lambda_b\to \Lambda_c \ell ^ - \overline{\nu }$  in  Eq.~\eqref{UD}. 
In particular,  $\alpha$ are found to be 
\begin{equation}\label{old}
\alpha(\Lambda_c \to p K^- \pi^+ , \Lambda l^+\nu) = (0.89\pm 0.10, -0.32),
\end{equation}
from the $SU(3)_F$ analysis~\cite{UP-down} and   light-front quark model~\cite{lightfront}, respectively. 

The  azimuthal angles are closely related to the triple product asymmetries,  which flip signs under TR transformation~\cite{TimeReversal}. 
To probe them, we define 
\begin{eqnarray}\label{Time}
&&{\cal T}_\ell  = \frac{1}{P_b }\left(  \int ^\pi _0 - \int_\pi  ^{2\pi } \right) \Gamma_{\phi_\ell}  d \phi_\ell  = - \frac{\pi^2 }{8\Gamma } \int^{(M_b -M_c)^2}_{m_\ell^2}   \zeta (q^2){\bf \text{Im}}( {\cal X}_7 )  dq^2 \,,\\
&&{\cal T}_c  = \frac{1}{\alpha  }\left[ \left(  \int ^\pi _0 - \int_\pi  ^{2\pi } \right)d \Phi \right]  \left[ \left(  \int _0^{1}  - \int^0   _{-1 } \right)d \cos \theta_c   \right]   \Gamma_{\phi_\ell, \cos \theta_c }   \nonumber\\
&&\quad =- \frac{2}{3 \pi \Gamma } \int^{(M_b -M_c)^2}_{m_\ell^2}   \zeta (q^2){\bf \text{Im}}( {\cal X}_{12} )  dq^2 \nonumber\,,
\end{eqnarray}
which are  proportional to the complex phases of $\xi_\pm$, and vanish without NP. 
Comparing to the direct {\it CP} asymmetries,  TR asymmetries do not require strong phases, which are  great advantages to probe {\it CP} violation as strong phases are absent in the semileptonic decays.
Note that one can also construct other TR asymmetries from ${\cal X}_{8,13,18\text{--}26}$.

\section{Contributions from possible new physics}
Let us 
consider the  dimension-six effective Hamiltonian from NP with left-handed neutrinos, read as 
\begin{align}
	\begin{aligned}
{\cal H}_{eff}^N = \frac{	G_F} {\sqrt{2}}V_{cb}&[ 
\overline{c}(C_S+C_P {\gamma_5})b  \left( \overline{u}_{\ell}   P_L  v   \right) +
\\
&     \overline{c}\gamma^{\mu} \left( C_L P_L+ C_R  P_R  \right) b \left(  \overline{u}_{\ell} {\gamma_{\mu}} P_L  v  \right)  +  C_T\left( \overline{u}_{\ell} \sigma_{\mu \nu}P_L v   \right) \overline{c}\sigma^ {\mu \nu}b ]\,,
	\end{aligned}
\end{align}
where $P_{R,L} = (1 \pm \gamma_5)$,  $C_{S,P,R,L,T}$ are the Wilson coefficients, which are complex and  depend on the lepton flavors in general, and $N$ in the superscript  indicates  that NP is considered. 
The effects of  $C_{S,P,R,L}$ can be absorbed by redefining the amplitudes as 
\begin{eqnarray}
\label{Right}
&&	a_\pm^N = \left(1 +C_L\right) a_\pm + C_R a_\mp \,,~~~b^N _\pm   =  \left(1 +C_L\right) b _ \pm + C_R b_\mp  \,,\nonumber\\
&&t_\pm ^{N} =\left(1 +C_L\right)  t_\pm +C_Rt_\mp  -  \frac{\sqrt{ Q_+q^2}}{m_\ell}C_S f_s \pm    \frac{\sqrt{ Q_- q^2}}{m_\ell} C_P g _p \,,
\end{eqnarray}
where 
$f_s$ and $g_p$ are defined by
\begin{equation}
 \langle  \Lambda_c | 
 \overline{c} \left( 1 +\gamma_5\right) b  | \Lambda_b\rangle = \overline{u}_c\left(
 f_s+ g_p \gamma_5 
 \right)u_b\,. 
\end{equation}
The derivations can be found in Appendix B. 
Note that in Eq.~\eqref{Right}, $\xi_\pm$ are calculated within the SM given in Eq.~\eqref{helicity}.  
The angular distributions can be easily obtained by substituting $\xi_\pm^N$  for $\xi_\pm$ in Table~\ref{Table}\,.
In the case of  $C_{R,S,P}=0$, the effect of  $C_L$   can be absorbed by  redefining  $V_{cb}$ as $V_{cb}(1+C_L)$, leaving
 the  angular distributions  unaltered. Therefore, in the following, we would simply take $C_L =0 $.

Let us first consider the case that $C_R\neq0 $ with $C_{S,P}=0$. 
For the total decay widths, 
$C_R$ would be polluted by the uncertainties of the form factors. 
However, we can utilize  that $\xi_\pm $ are real, whereas $C_R$ can be complex in general.  Plugging  Eq.~\eqref{Right} in Eq.~\eqref{equationTR}, we arrive at
\begin{eqnarray}\label{New1}
{\bf \text{Im}} \left( {\cal X}_7^{ N} \right) = 3\sqrt{2}\,{\bf \text{Im}}(C_R) \left(
a_+ b_+ - a_- b_- 
\right)\,, \nonumber\\
{\bf \text{Im}} \left( {\cal X}_{12} ^{ N} \right) = 3\sqrt{2}\,{\bf \text{Im}}(C_R) \left(
a_+ b_- - a_- b_+ 
\right) \,,
\end{eqnarray}
where we have taken $\xi _\pm$ as real, calculated by Eq.~\eqref{helicity}.  

On the other hand,
the effects of $C_{S,P}$ are largely enhanced by the smallness of the lepton quark masses when $q^2/m_\ell^2 \gg 1 $.  Therefore, measuring $t_\pm$ in high $q^2$ regions would be useful to constrain the values  of $C_{S,P}$.  
To diminish the uncertainties from the form factors, one can examine the 
 complex phases, given by 
\begin{eqnarray}\label{New2}
&&{\bf \text{Im}} \left( {\cal X}_{7}^{ N} \right) =  \frac{3}{ \sqrt{2}}m_\ell \left[ - 
{\bf \text{Im}} (C_S) \sqrt{ \frac{Q_+ }{q^2}}  f_s \left(
b_+ + b_-
\right)
+  {\bf \text{Im}} (C_P) \sqrt{ \frac{Q_- }{q^2}}  g_p \left(
b_- -  b_+
\right)
\right]\,,\nonumber\\
&&{\bf \text{Im}} \left( {\cal X}_{12}^{ N} \right)  = \frac{3}{ \sqrt{2}}m_\ell \left[
{\bf \text{Im}} (C_S)  \sqrt{ \frac{Q_+ }{q^2}}    f_s \left(
b_+ + b_-
\right)
+  {\bf \text{Im}} (C_P) \sqrt{ \frac{Q_- }{q^2}} g_p  \left(
b_- -  b_+
\right)
\right]\,,
\end{eqnarray} 
where we have taken $C_R=0$. 
By collecting Eqs.~\eqref{New1} and \eqref{New2},
the net effects of NP on ${\cal T}_{\ell, c }$  are summarized as  follows
\begin{eqnarray}\label{Newph}
\mathcal{T}_{\ell}&=&-\frac{3 \pi^2}{8 \sqrt{2}}\left(\operatorname{Im}\left(C_R\right) \mathcal{Y}_R-\operatorname{Im}\left(C_S\right) \mathcal{Y}_S+\operatorname{Im}\left(C_P\right) \mathcal{Y}_P\right)\nonumber\\
{\cal T}_c &=&
-\frac{\sqrt{2}}{\pi }\left(  
{\bf \text{Im}}(C_R) {\cal Y}'_R +   {\bf \text{Im}}(C_S) {\cal Y}_S + {\bf \text{Im}}(C_P) {\cal Y}_P\right) \,,
\end{eqnarray}
where 
\begin{eqnarray}
{\cal Y}_R^{(\prime)}   &=&  \frac{1 }{ \Gamma } \int^{(M_b -M_c)^2}_{m_\ell^2}2 \zeta \left(  a_+ b_{+(-)} - a_- b_{-(+)} \right)dq^2  \,,\nonumber\\
{\cal Y}_S  &=&  \frac{1 }{ \Gamma } \int^{(M_b -M_c)^2}_{m_\ell^2}\zeta  m_\ell \sqrt{ \frac{Q_+ }{q^2}}   f_s\left( b_+ + b_-  \right)dq^2\,,\nonumber\\
{\cal Y}_P  &=&  \frac{1 }{ \Gamma } \int^{(M_b -M_c)^2}_{m_\ell^2}\zeta m_\ell \sqrt{ \frac{Q_+ }{q^2}}   g_p \left( b_- -  b_+   \right)dq^2\,.
\end{eqnarray}
Notice that $\Gamma$ also depends on $C_{R,S,P}$. However, in this work, we take $C_{R,S,P}$ as zero in $\Gamma$ as a first order approximation, and therefore ${\cal Y}^{(\prime)}_{R,S,P}$  can be computed once the form factors are given. 

To examine TR asymmetries in the experiments, we define 
\begin{eqnarray}\label{texp}
\Delta N_{\ell}  \equiv &&
	N(\pi> \phi_{\ell} >0) -  N(2\pi> \phi_{\ell} > \pi)  = \epsilon N_{\Lambda_b}  {\cal T}_\ell  {\cal B} ( \Lambda_b \to \Lambda_c^+ \ell^- \nu)
	P_b
	\,,\nonumber\\
\Delta N _{f }  \equiv &&
N(\pi> \phi_{c} >0, \cos \theta_c >0 )  +  N(2\pi> \phi_{c} > \pi,\cos \theta_c <0 )\nonumber  \\
&&~~~~-N(\pi> \phi_{c} >0, \cos \theta_c < 0 )  -  N(2\pi> \phi_{c} > \pi,\cos \theta_c > 0 ) \nonumber\\
= &&\epsilon N_{\Lambda_b}  {\cal T}_c  {\cal B} ( \Lambda_b \to \Lambda_c^+ \ell^- \nu)    {\cal B}( \Lambda_c \to B_n f )
\alpha( \Lambda_c \to B_n f )
\,,
\end{eqnarray}
which hold at ${N\to \infty}$ with  $N$  the  number of the observed events, where $N_{\Lambda_b}$ is the numbers of $\Lambda_b$ in  experiments,  and $\epsilon$ is the efficiency for the experimental reconstruction. 
To reduce the statistical uncertainties in $\Delta N_f$, we can sum over the decay modes of $\Lambda_c^+$, given as 
\begin{equation}
\Delta N _c = \sum _f \left|
N_f  
\right| = \epsilon N_{\Lambda_b}  {\cal T}_c  {\cal B} ( \Lambda_b \to \Lambda_c^+ \ell^- \nu)   \sum_{B_nf} \left|  {\cal B}( \Lambda_c \to B_n f )
\alpha( \Lambda_c \to B_n f )\right| \,. 
\end{equation}
As $\Delta N_{{\ell, c }}$ are proportional to ${\cal T}_{\ell,c} $, a nonzero value of 
$\Delta N_{{\ell }}$ or $\Delta N_{{ c }}$
 would be a smoking gun  of NP. 

 The  full angular distributions including  the tensor operator are given in Appendix~C.  
 For simplicity,
 we take $C_T=0$ in the numerical analysis, 
as  they can not be reduced to  the form of  Eq.~\eqref{good}, which  breaks the  angular analysis. In addition, 
the tensor operator is closely related to the scalar ones  by the Fierz transformation in the leptoquark effective field theory~\cite{Leptoquark}.

\section{Numerical results}
As mentioned in Sec.~\MakeUppercase{\romannumeral 3},  nonzero  signals of ${\cal T}_{\ell,c}$  can be clear evidence of
{\it CP} violation from NP with $t^N_\pm$  largely enhanced  in  high $q^2$ regions by $C_{S,P}$. 
In Eq.~\eqref{Newph}, ${\bf \text{Im}}(C_{R,S,P})$ are in general free parameters of NP, whereas ${\cal Y}_{R,S,P}^{(\prime)}$ can be computed by the form factors.
In this work, we utilize the homogeneous bag model to estimate  the form factors~\cite{Liu:2022pdk}, which agree well with those by  the lattice QCD calculation
as well as the heavy quark symmetry~\cite{lattice}.

The values of  ${\cal Y}_{R,S,P}^{(\prime)}$ are  listed in Table~\ref{NewTable}, from which one can see that ${\cal Y}_R$ and ${\cal Y}_R'$ are both sizable for all  flavors, giving us a good opportunity to examine Im$(C_R)$. 
In contrast, the values of 
${\cal Y}_{S,P}$ for $\ell = e$ and $\mu$
 are suppressed due to $m_\ell$. For $\ell = \tau $, ${\cal Y}_{S,P}$ are still 3 times smaller than ${\cal Y}_R^{(\prime)}$.
Hence, Im$(C_{S,P})$ are  much more difficult  to be observed  in the experiments  comparing to Im$(C_R)$.

\begin{table}[b]
	
	\caption{Parameters defined  in Eq.~\eqref{Newph}, where the uncertainties come  from the model calculation. }
	\label{NewTable}
	\begin{tabular}{lcccccc}
		\hline
		\hline
		$\ell$ &		${\cal Y}_R$ &		${\cal Y}_R' $ &		${\cal Y}_S$  &		${\cal Y}_P$   \\
		\hline
		$e$ &$-0.238(11)$ &$ 0.703(13)$ & $<10^{-4}$ & $<10^{-4} $\\
		$\mu$ &$-0.237(12)$ &$ 0.701(13)$& $-0.0037(2)$ &$ 0.0136(2)$ \\
		$\tau $&$-0.149(6)$ &$ 0.438(8)$ & $-0.039(2)$  & $ 0.180(2)$ \\
		\hline
		\hline
		
	\end{tabular}
\end{table}

\begin{figure}[t]
	\includegraphics[width=0.6 \linewidth]{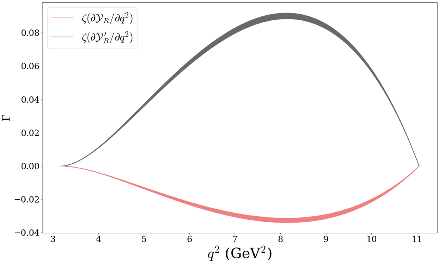}  	
	\caption{ $q^2$ dependence of   $\zeta (\partial {\cal Y}_{R}^{(\prime)}/\partial q^2) $ in  $\Lambda_b \to \Lambda_c \tau^-\overline{\nu}$, where the bands represent the uncertainties caused by the form factors.   }
	\label{CR} 
\end{figure}

  To explain the excesses of $R_{D^{(*)}}$,  $C_{R,S,P}$ are found to be tiny  for $\ell =e $ and $\mu$, but
fortunately, $C_R = \pm 0.42 (7) i$ is huge for $\ell = \tau $~\cite{Leptoquark}. 
We have plotted $\zeta ( \partial {\cal Y}_{R}^{(\prime)}/ \partial q^2 )$ for $\ell = \tau$ in FIG.~\ref{CR}, where the bands represent the uncertainties from the form factors. 
One can see that the ideal $q^2$ region to search for  the asymmetries lies  around $7$~GeV$^2 <q^2 <9$~GeV$^2$,  since  they are huge within the region.  
Finally, putting  the values of ${\cal Y}_R^{(\prime)}$  and $C_R=\pm 0.42(7)i$  in Eq.~\eqref{Newph},
we find   that 
\begin{equation}
{\cal T}_{\ell} = \pm 0.16(3)\,,~~~~{\cal T}_{c } = \pm 0.08(2)\,,
\end{equation}
 for $\Lambda_b \to \Lambda_c \tau^- \overline{\nu}$.
 Notice that the signs are irrelevant for searching evidence of  NP as long as they are nonzero.
 To estimate the results   at LHCb run2, we take
 $N_{\Lambda_b} = 5 \times 10 ^9$,  $P_b=0.03$, $\varepsilon = 10^{-4}$, and 
\begin{equation}\label{40}
\sum_f \left|  {\cal B}( \Lambda_c \to B_n f )
\alpha( \Lambda_c \to B_n f )\right|  =  6\times 10^{-2} \,,
\end{equation}
resulting in that
$| \Delta N_\ell | \approx  50$  and  $\Delta N _c  \approx  20$ for $ \ell = \tau$,
which are large and ready to be measured. 
Here, Eq.~\eqref{40} is derived by crunching up the numbers in Eqs.~\eqref{alphap} and \eqref{old}.

\begin{figure}[h]
	\includegraphics[width=0.6 \linewidth]{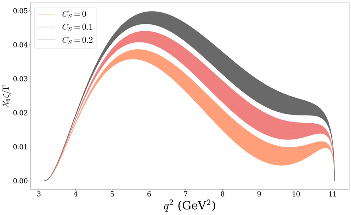}  	
	\caption{ $q^2$ dependence of ${\cal X}_{6}\zeta/\Gamma$ with $\ell = \tau$ for $C_S=0,0.1$ and $0.2$, respectively.    }
	\label{X16} 
\end{figure}

To probe the effects of the scalar operators, we find that 
${\cal X}_6$, which can be understood as a combination of 
$A_{FB}$ and $A_{UD}$, is sensitive to $C_{S,P}$ for $\ell = \tau$. 
The results are plotted in FIG.~\ref{X16}, where we have taken  $C_P=C_S$\footnote{The scenario of $C_P=-C_S$  is ruled out by the lifetime of $B_c^-$~\cite{Bc}. 
}.
In the region of 9 GeV$^2< q^2<$10 GeV$^2$,  ${\cal X}_6$  can be enhanced largely.  In particular, it is  twice larger with  $C_S=0.2$ in comparison to  that in the SM.

\section{Summary}
Based on the helicity formalism, we have given the full angular distributions of $\Lambda_b \to \Lambda_c(\to B_n f ) \ell^- \overline{\nu}$. In particular, we have  identified TR violating 
terms, which vanish in the SM due to the lack of relative complex phases. 
Since strong phases are not required in these TR violating observables in contrast to the direct {\it CP} asymmetries, they can be reliably calculated.
The angular distributions have been  given explicitly with the helicity amplitudes in Table~\ref{Table}.  We have cross-checked our  results with those in  Refs.~\cite{Korner,KornerNeutron}, and found that they are consistent. 
Note that our results can be easily applied to $\Xi_b\to\Xi_c (\to B_nf)\ell^-\overline{\nu}$ with trivial modifications.

Notably, the effects of NP can be absorbed by redefining the helicity amplitudes as demonstrated in Eq.~\eqref{Right} with $\xi_\pm$ calculated in the SM. 
We recommend the experiments to measure the TR violating observables of
 ${\cal T}_{\ell,c}$ defined in Eq.~\eqref{Time} for searching NP as they vanish in the SM.  To compare with the experiments,  $\Delta N_{\ell, c}$ have been defined by  the  numbers of the observed events, which are proportional to ${\cal T}_{\ell,c}$.  Based on  $C_R= \pm 0.42(7)i$ for $\ell = \tau$, we have  obtained that $| \Delta N_{\ell, c} | \approx  50, 20 $ at LHCb run2, which are sufficient for   measurements. On the other hand, we have pointed out that ${\cal X}_6$ is sensitive to $C_{S,P}$ for $\Lambda_b\to \Lambda_c \tau^- \overline{\nu}$, which can be largely enhanced in the  high $q^2$ region.

\begin{acknowledgments}
	We would like to thank Jiabao Zhang for  valuable discussions.
	This work is supported in part by the National Key Research and Development Program of China under Grant No. 2020YFC2201501 and  the National Natural Science Foundation of China (NSFC) under Grant No. 12147103.
\end{acknowledgments}

\appendix
\section{Dirac spinors}
In  this work, we  choose
  fermions and  antifermions in $p\hat{z}$ and $-p\hat{z}$ directions, respectively.   We have  that
\begin{equation}
	u_+ = 
	\left(
	\begin{array}{c}
		E_+\\ 
		0\\
		E_-\\
		0
	\end{array}
	\right) \,, \quad  
	u_- = 
	\left(
	\begin{array}{c}
		0\\ 
		E_+\\
		0\\
		-E_- 
	\end{array}
	\right) \,, \quad  
	v_+ = 
	\left(
	\begin{array}{c}
		E_-\\
		0\\
		-E_+\\
		0 
	\end{array}
	\right) \,,\quad  
	v_- = 
	\left(
	\begin{array}{c}
		0\\ 
		-E_-\\
		0\\
		-E_+ 
	\end{array}
	\right) \,, 
\end{equation}
with  $`` \pm " $  denoting the helicities,  $E_\pm = \sqrt{E\pm m}$, and $m$ the particle mass. 
Notice that the relative signs   are crucial,  
  fixed by the relations
\begin{equation}
u_- = L^u_z R_y (\pi) (L_z^{u})^{-1} u _+\,,~~~v_- = (L_z^{v})^{-1} R_y (\pi) L_z^v v_+\,,
\end{equation}
where $R_y$ is the rotation matrix  toward $\hat{y}$, and 
$L_z^{u,v}$ are  the Lorentz boost  operators toward $\hat{z}$. Here, $L_z^{u,v}$ are taken in a way such that $L_z^{u-1}u_\pm$ and $L_z^v v_\pm$ are at rest. 
\section{Angular distributions in the standard model}\label{APPB}
We now sketch the derivation of Eq.~\eqref{distru}. We start with a two-body decay of $i\to f_1f_2$, where $i$ and $f_{1,2}$ are unspecific particles. The decay distribution is given as 
\begin{eqnarray}\label{B1}
	&& \frac{\partial^2 \Gamma (i\to f_1f_2)}{\partial \phi \partial \cos \theta}
	\propto \sum_{\lambda_1,\lambda_2} \left|\langle
p, \theta, \phi,\lambda_1,\lambda_2 | U(\infty , -\infty ) | i; J , J_z\rangle 
\right| ^2 \,,\\
&&| p,\theta, \phi,\lambda_1,\lambda_2  \rangle  = R_z(\phi) R_y(\theta ) \left( | f_1; \vec{p}= p \vec{z} , \lambda_1\rangle \otimes 
| f_2; \vec{p}= - p \vec{z} , \lambda_2\rangle \right) \,,
\end{eqnarray}
where $U$ is the time evolution operator,  $J$ and $J_z$ are the angular momentum and its $z$ component of the initial particle, $R_{y,z}$ are the rotational operators pointing toward $(y,z)$, and  $\lambda_{1,2}$ are the helicities of $f_{1,2}$. 
In the decay distributions, we have to sum over the helicities of the outgoing particles as they are difficult to be probed in the experiments. 

In  two-body  systems,  states with definite angular momenta and helicities can be constructed  as 
\begin{eqnarray}\label{HelicityStates}
	|\lambda_1\,, \lambda_2 , J, J_z \rangle = 
	\frac{1}{(2J+1 )\pi}
	\int d\cos \theta d \phi  |p, \theta , \phi , \lambda_1\,, \lambda_2\rangle  e^{iJ_z\phi} d^J(\theta) ^{J_z}\,_{\lambda_1 - \lambda_2}\,,
\end{eqnarray}
along with the identity
\begin{equation}\label{B3}
	1 = \sum_{J,J_z}  \frac{4\pi }{2 J+1 }| J , J_z, \lambda_1 , \lambda_2 \rangle \langle  J , J_z, \lambda_1 , \lambda_2 |.
\end{equation}
Notice that in Eq.~\eqref{HelicityStates}, $\lambda_{1,2}$ are unaltered because  they are rotational scalars. By inserting  Eq.~\eqref{B3} in Eq.~\eqref{B1}, we obtain
\begin{equation}\label{B5}
\frac{\partial^2 \Gamma (i\to f_1f_2)}{\partial \phi \partial \cos \theta} \propto \sum_{\lambda_1,\lambda_2}  \left|
e^{iJ_z\phi} d^{J}(\theta) ^{J_z} \,_{\lambda_1 - \lambda_2}  H_{\lambda_1\lambda_2} 
\right|^2 \,,
\end{equation}
with 
\begin{equation}
H_{\lambda_1,\lambda_2 } \equiv \langle J, J_z,\lambda_1,\lambda_2|  U(\infty , -\infty ) | i ; J , J_z\rangle \,.
\end{equation}
Clearly, $H_{\lambda_1,\lambda_2 }$ is  independent of $J_z$ since $U$ must be a scalar. 
In Eq.~\eqref{B5}, we see that the decay distributions are separated into two different parts. The kinematic part is described by the Wigner-d matrix whereas the dynamical part  $H_{\lambda_1,\lambda_2}$. 

The three-body decay distributions can be obtained by decomposing the systems into a product of two-body decays as demonstrated in Eq.~\eqref{good}. 

\section{Contributions from scalar and current operators}
The contributions of $C_L$  are  given by 
\begin{equation}\label{app0}
\xi_\pm \to (1 +C_L) \xi_\pm\,,
\end{equation}
 due to  the same coupling in the SM. Furthermore, those of $C_R$ can be obtained  straightforwardly by 
\begin{equation}\label{app1}
\xi_\pm \to \xi_\pm + C_R \xi_\mp\,,
\end{equation}  
as $\overline{c} \gamma^\mu P_{L,R} b$ are related by the parity.

The scalar operators contribute to the amplitudes as 
\begin{eqnarray}
&&\frac{G_f}{\sqrt{2}} V_{cb}  L_t^N B_t^N\,,~~~L_t^N =  {\cal C} \overline{u}_{\ell}  P_L v_{\nu}  \,,~~~B_t^N=  {\cal C}^{-1} \overline{u}_c\left(C_S
f_s+C_P  g_p \gamma_5 
\right)u_b \,,
\end{eqnarray}
where ${\cal C}$ is a  constant,
Clearly, $B_t^N$ and $L_t^N$ can be viewed  as the transitions of  $\Lambda_b \to \Lambda_c P^-$ and   $P^-\to l^-\overline{\nu}$, respectively, with $P^-$ an effective particle from NP.  As $P^-$ is spinless, $L_t^NB_t^N$ is only
  related to $L_tB_t$ in Eq.~\eqref{good}. 
By adjusting ${\cal C}$ such that $L_t^N = L_t$, we arrive  at
\begin{equation}\label{app2}
t_\pm ^{N} = t_\pm  -  \frac{\sqrt{ Q_+q^2}}{m_\ell}C_S f_s \pm   \frac{\sqrt{ Q_- q^2}}{m_\ell} C_P g _p \,.
\end{equation}
By collecting Eqs.~\eqref{app0}, \eqref{app1} and \eqref{app2}, we obtain Eq.~\eqref{Right}.
It  is interesting to see that all the contributions can be encapsulated  in $\xi_\pm$, which already exist in the  SM.

\section{Contributions from  tensor operator}
To cooperate the tensor operator  with the helicity amplitudes, by utilizing Eq.~\eqref{eq2} we decompose the products of the Minkowski metric as 
\begin{eqnarray}
	\left(  
	g^{\mu\nu}	g^{\mu'\nu'} - 	g^{\mu'\nu}	g^{\mu\nu'}\right) = \sum_\lambda \left(-   V_{1}^{\mu \mu'}(\lambda) V_1^{\nu \nu' *} (\lambda) +  V_2^{\mu \mu'} (\lambda)V_2 ^{\nu \nu' *} (\lambda) \right) \,,
\end{eqnarray}
with 
\begin{eqnarray}
V_1^{\mu \mu'}(\lambda)   = \left( \varepsilon_t^\mu \varepsilon^{\mu'}_\lambda -
	\varepsilon_t^{\mu'} \varepsilon^{\mu}_\lambda 
	\right)  \,, ~~~
V_2^{\mu \mu'}(\lambda)  = \frac{1}{\sqrt{2}}\left( \varepsilon_{\lambda_1}^\mu \varepsilon^{\mu'}_{\lambda_2} -
	\varepsilon_{\lambda_1}^{\mu'} \varepsilon^{\mu}_{\lambda_2}
	\right) \,,
\end{eqnarray}
and 
\begin{equation}
( \lambda_1,\lambda_2 ) = (1,0), (1,-1), (0,-1)\,,~~\text{for} ~~\lambda = 1,0, -1\,. 
\end{equation}
To see that $V_{1,2}$ can be viewed as  spacelike vectors,
in the $\vec{q}$ frame, we  define 
\begin{equation}\label{spin1 }
\left (\vec{V}_1\right )_{i}  =  V_1 ^{0 i} \,,~~~\left (\vec{V}_2\right )_{i} = \epsilon_{ijk}\vec{V}_2 ^{jk} \,,
\end{equation}
with $i,j,k = x,y ,z$ and $\epsilon_{ijk}$ the totally antisymmetric tensor. 
It shows that $V_1$ and $V_2$ are spacelike, which are essentially spin-1  under the  $SO(3)$ rotational group. 

The results can be understood in terms of the group theory, given  as\footnote{In the $SO(3)$  group, $\overline{{\bf 3 }}$  and ${\bf 3}$ are equivalent. However, they behave differently under  the $O(3)$ group, where $\overline{ {\bf 3}}$ and ${\bf 3}$ are parity even and odd, respectively. }
\begin{equation}\label{group}
\left( {\bf 1} \oplus {\bf 3}\right)_1  \otimes \left( {\bf 1} \oplus {\bf 3}\right)_2   = \left(
{\bf 1 } \oplus  {\bf 3}  \oplus {\bf 6} 
\right)_S + \left(  {\bf 3}   \oplus  \overline{ 
{\bf 3 } } 
\right)_A\,,
\end{equation}
 where ${\bf 1}$, ${\bf 3}$, and  ${\bf 6}$ are the representations of the $SO(3)$ rotational group, and we have used the fact that a four-vector 
 is ${\bf 1}\oplus {\bf 3}$.
 In Eq.~\eqref{group}, $`` S" $ and $`` A " $ in the subscripts indicate symmetric and antisymmetric between the first and  second objects, respectively.   
  The antisymmetric nature of $\sigma_{\mu \nu}$ forces us to select 
the second solution, where  ${\bf 3}$ and $\overline{{\bf 3}}$  correspond to $V_1$ and $V_2$, respectively.

Now, we are able to rewrite 
the transition matrix element of the tensor operator as 
\begin{eqnarray}
&&	 g^{\mu \nu}g^{\mu ' \nu'} \left( \overline{u}_{\ell} \sigma_{\mu \mu'}P_L v   \right)
\langle \Lambda_c \left|
 \overline{c}\sigma_ {\nu \nu' }b\right|\Lambda_b \rangle= \sum _{\lambda , n }	L^{V_n}_\lambda  B_\lambda ^{V_n} 
 \,,\nonumber\\
	&&	L^{V_n}_\lambda = 	\frac{(-1)^{n} }{2} i V_n^{\mu \mu'} (\lambda) \overline{u}_{\ell} \sigma_{\mu \mu'}P_L v    \,,\qquad  B_\lambda ^{V_n}  =  - i 
   V_n^{\nu \nu' *} (\lambda) \langle \Lambda_c \left|
   \overline{c}\sigma_ {\nu \nu' }b\right|\Lambda_b \rangle \,,
\end{eqnarray}
which  can be understood as a product of $\Lambda_b \to \Lambda_c V_n$ and $V_n \to \ell ^-\overline{\nu}$, with  $V_n$ effectively spin-1 particles. The helicity amplitudes of the lepton sector are given as  
\begin{eqnarray}
\left(  h^{V_1} _+ , h^{V_1} _- \right)  =\left(  -  \frac{1}{ \sqrt{2}}  h_-\,,-\sqrt{2} h_+\right) \,,~~~ \left( h^{V_2} _+ , h^{V_2 } _- \right) =\left(  \frac{1}{2}h_-\,,  h_+\right) \,,
\end{eqnarray}
with 
\begin{equation}
h_\pm^{V_{1,2}} = L^{V_{1,2} }_{\pm \frac{1}{2}  - \frac{1}{2}  }  \left(\lambda_{\ell} = \pm \frac{1}{2}, \vec{p}_\ell = - \vec{p}_\nu = p \hat{z}\right)  \,,
\end{equation}
 describing
 $( V_1 \to \ell^- \overline{\nu}) $ and $( V_2  \to \ell^- \overline{\nu}) $, respectively. 
For the baryon sector, the helicity amplitudes read as 
\begin{equation}
H_{\lambda_c, \lambda}  ^{V_{n} } =   B^{V_n}_{\lambda} \left(
\lambda_b = \lambda_c - \lambda , \vec{p}_c = - \vec{q} = |\vec{p}_c| \hat{z}
\right)\,,
\end{equation}
 with $H_{\lambda_c, t}^{V_n}= 0 $ and $V_{1,2}$ being spacelike. 

The sixfold angular distributions now take the form 
\begin{eqnarray}\label{distruT}
&&{\cal B}(\Lambda_c \to B_n f)\zeta(q^2)\sum_{\lambda_\ell \,,\lambda \,,\lambda_b  }\rho_{\lambda_b,\lambda_b }\left| A^c_{\lambda }	\sum_{\lambda_c\,, \lambda_W } 
\left[ C_T h_{ \lambda_\ell } ^{V_1} \left(  H_{\lambda_c,\lambda_W}^{V_1} - \frac{1}{\sqrt{2} } H_{\lambda_c,\lambda_W}^{V_2}
\right) 
 \right. \right.  \\
&& \left.  + (-1) ^{J_W  } H_{\lambda_c\,,\lambda_W }  h_{ \lambda_\ell }  \bigg]  
	d^{\frac{1}{2}}(\theta_b)^{\lambda_b }\, _{\lambda_c - \lambda_W }
	d^{\frac{1}{2}}(\theta_c)^{\lambda_c}\,_{\lambda }
	d^{J_W }(\theta_\ell )^{\lambda_W }\,_{\lambda_\ell  - \frac{1
		}{2}}
	e^{i(\lambda_c\phi_c +\lambda_\ell  \phi_\ell )} 
	\right|^2\,, \nonumber
\end{eqnarray}
which  cannot be reduced to Eq.~\eqref{distru} by redefining the amplitudes. Thus, Table~\ref{Table} would no longer be suitable after the tensor operator is considered.

\end{document}